\input harvmac
\noblackbox
\let\includefigures=\iftrue
\let\useblackboard=\iftrue
\newfam\black

%Figure Stuff
\includefigures
\message{If you do not have epsf.tex (to include figures),}
\message{change the option at the top of the tex file.}
\input epsf
\def\figin{\epsfcheck\figin}\def\figins{\epsfcheck\figins}
\def\epsfcheck{\ifx\epsfbox\UnDeFiNeD
\message{(NO epsf.tex, FIGURES WILL BE IGNORED)}
\gdef\figin##1{\vskip2in}\gdef\figins##1{\hskip.5in}% blank space instead
\else\message{(FIGURES WILL BE INCLUDED)}%
\gdef\figin##1{##1}\gdef\figins##1{##1}\fi}
\def\DefWarn#1{}
\def\figinsert{\goodbreak\midinsert}
\def\ifig#1#2#3{\DefWarn#1\xdef#1{fig.~\the\figno}
\writedef{#1\leftbracket fig.\noexpand~\the\figno}%
\figinsert\figin{\centerline{#3}}\medskip\centerline{\vbox{
\baselineskip12pt\advance\hsize by -1truein
\noindent\footnotefont{\bf Fig.~\the\figno:} #2}}
\bigskip\endinsert\global\advance\figno by1}
%%%
\else
\def\ifig#1#2#3{\xdef#1{fig.~\the\figno}
\writedef{#1\leftbracket fig.\noexpand~\the\figno}%
%\figinsert\figin{\centerline{#3}}\medskip
%\centerline{\vbox{\baselineskip12pt
%\advance\hsize by -1truein\noindent
%\footnotefont{\bf Fig.~\the\figno:} #2}}
%\bigskip\endinsert
\global\advance\figno by1}
\fi
%

%%BLACKBOARD FONT STUFF
\useblackboard
\message{If you do not have msbm (blackboard bold) fonts,}
\message{change the option at the top of the tex file.}
\font\blackboard=msbm10 scaled \magstep1
\font\blackboards=msbm7
\font\blackboardss=msbm5
\textfont\black=\blackboard
\scriptfont\black=\blackboards
\scriptscriptfont\black=\blackboardss

\else

\fi
% *************************************
%\draft
%
\def\subsubsec#1{\bigskip\noindent{\it{#1}} \bigskip}
\def\yboxit#1#2{\vbox{\hrule height #1 \hbox{\vrule width #1
\vbox{#2}\vrule width #1 }\hrule height #1 }}
\def\fillbox#1{\hbox to #1{\vbox to #1{\vfil}\hfil}}
\def\ybox{{\lower 1.3pt \yboxit{0.4pt}{\fillbox{8pt}}\hskip-0.2pt}}
%
%
%%MATH MACROS
%Greek letters and their bars

%More bars

\def\comments#1{}

\def\half{{1\over 2}}

\def\tr{{\rm tr\ }}

\def\CC{{\cal C}}
\def\CD{{\cal D}}

\def\CF{{\cal F}}

\def\CN{{\cal N}}
%AEL

%AEL

\def\II{\relax{I\kern-.10em I}}

\def\IZ{\relax\ifmmode\mathchoice
{\hbox{\cmss Z\kern-.4em Z}}{\hbox{\cmss Z\kern-.4em Z}}
{\lower.9pt\hbox{\cmsss Z\kern-.4em Z}}
{\lower1.2pt\hbox{\cmsss Z\kern-.4em Z}}
\else{\cmss Z\kern-.4emZ}\fi}
\def\IB{\relax{\rm I\kern-.18em B}}
\def\IC{{\relax\hbox{$\inbar\kern-.3em{\rm C}$}}}
\def\ID{\relax{\rm I\kern-.18em D}}
\def\IE{\relax{\rm I\kern-.18em E}}
\def\IF{\relax{\rm I\kern-.18em F}}
\def\IG{\relax\hbox{$\inbar\kern-.3em{\rm G}$}}
\def\IGa{\relax\hbox{${\rm I}\kern-.18em\Gamma$}}
\def\IH{\relax{\rm I\kern-.18em H}}
\def\II{\relax{\rm I\kern-.18em I}}
\def\IK{\relax{\rm I\kern-.18em K}}
\def\IP{\relax{\rm I\kern-.18em P}}
%\def\IX{\relax{\rm X\kern-.01em X}}
%this doesn't work

%

\def\inbar{\,\vrule height1.5ex width.4pt depth0pt}

\font\cmss=cmss10 
\def\IR{\relax{\rm I\kern-.18em R}}

%

 % for now

%

\def\lp10{\ell_p^{10}}
\def\lp11{\ell_p^{11}}
\def\R11{R_{11}}

\def\frac#1#2{{#1 \over #2}}

%%ENGLISH MACROS
\def\eg{{\it e.g.}}
\def\ie{{\it i.e.}}

\hyphenation{Di-men-sion-al}

%%REFERENCING MACROS

%%

%\OoguriGX
\lref\OoguriGX{
H.~Ooguri and C.~Vafa,
``Worldsheet derivation of a large N duality,''
arXiv:hep-th/0205297.
%%CITATION = HEP-TH 0205297;%%
}

%\CachazoJY
\lref\CachazoJY{
F.~Cachazo, K.~A.~Intriligator and C.~Vafa,
``A large N duality via a geometric transition,''
Nucl.\ Phys.\ B {\bf 603}, 3 (2001)
[arXiv:hep-th/0103067].
%%CITATION = HEP-TH 0103067;%%
}

\lref\DijkgraafDH{R.~Dijkgraaf and C.~Vafa,
``A perturbative window into non-perturbative physics,''
arXiv:hep-th/0208048.
%%CITATION = HEP-TH 0208048;%%
}

\lref\DijkgraafVW{R.~Dijkgraaf and C.~Vafa,
``On geometry and matrix models,''
Nucl.\ Phys.\ B {\bf 644}, 21 (2002)
[arXiv:hep-th/0207106].
%%CITATION = HEP-TH 0207106;%%
}

\lref\DijkgraafFC{R.~Dijkgraaf and C.~Vafa,
``Matrix models, topological strings, and supersymmetric gauge theories,''
Nucl.\ Phys.\ B {\bf 644}, 3 (2002)
[arXiv:hep-th/0206255].
%%CITATION = HEP-TH 0206255;%%
}

\lref\KazakovCQ{V.~A.~Kazakov,
``A Simple Solvable Model Of Quantum Field Theory Of Open Strings,''
Phys.\ Lett.\ B {\bf 237}, 212 (1990).
%%CITATION = PHLTA,B237,212;%%
}

\lref\DistlerMT{J.~Distler and C.~Vafa,
``A Critical Matrix Model at c = 1,''
Mod.\ Phys.\ Lett.\ A {\bf 6}, 259 (1991).
%%CITATION = MPLAE,A6,259;%%
}

\lref\ArgyresFW{P.~C.~Argyres and A.~D.~Shapere,
``The Vacuum Structure of N=2 SuperQCD with Classical Gauge Groups,''
Nucl.\ Phys.\ B {\bf 461}, 437 (1996)
[arXiv:hep-th/9509175].
%%CITATION = HEP-TH 9509175;%%
}

\lref\DijkgraafPP{R.~Dijkgraaf, S.~Gukov, V.~A.~Kazakov and C.~Vafa,
``Perturbative Analysis of Gauged Matrix Models,''
arXiv:hep-th/0210238.
%%CITATION = HEP-TH 0210238;%%
}

\lref\BrezinSV{E.~Brezin, C.~Itzykson, G.~Parisi and J.~B.~Zuber,
``Planar Diagrams,''
Commun.\ Math.\ Phys.\  {\bf 59}, 35 (1978).
%%CITATION = CMPHA,59,35;%%
}

\lref\GopakumarKI{R.~Gopakumar and C.~Vafa,
``On the gauge theory/geometry correspondence,''
Adv.\ Theor.\ Math.\ Phys.\  {\bf 3}, 1415 (1999)
[arXiv:hep-th/9811131].
%%CITATION = HEP-TH 9811131;%%
}

\lref\Gorsky{
A.~Gorsky,
``Konishi anomaly and N=1 effective superpotentials from matrix models, ``
hep-th/0210281.
}

\lref\Ferretti{
R. Argurio, V. L. Campos, G. Ferretti, R. Heise,
``Exact Superpotentials for Theories with Flavors via a Matrix Integral,''
hep-th/0210291.}

\lref\BerensteinZW{
D.~Berenstein, E.~Gava, J.~M.~Maldacena, K.~S.~Narain and H.~Nastase,
``Open strings on plane waves and their Yang-Mills duals,''
arXiv:hep-th/0203249.
%%CITATION = HEP-TH 0203249;%%
}

\def\NPB#1#2#3{Nucl.~Phys~B{\bf#1} (#2) #3}
\def\PLB#1#2#3{Phys.~Lett.~B{\bf#1} (#2) #3}
\def\PRL#1#2#3{Phys.~Rev~Lett.~{\bf#1} (#2) #3}
\lref\fieldtheoryrefs{\eg\
%N. Seiberg and E. Witten, ``Electric-Magnetic Duality,
%        Monopole Condensation, and Confinement in N=2 Supersymmetric
%        Yang-Mills Theory,'' hep-th/9407087, \NPB{426}{1994}{19};
N. Seiberg and E. Witten, ``Monopoles, Duality, and Chiral
        Symmetry Breaking in N=2 Supersymmetric QCD,'' hep-th/9408099,
        \NPB{431}{1994}{484};
P.C. Argyres and A.E. Faraggi, ``Vacuum Structure and Spectrum
        of N=2 Supersymmetric SU(n) Gauge Theory,'' hep-th/9411057,
        \PRL{74}{1995}{3931};
A. Klemm, W. Lerche, S. Theisen and S. Yankielowicz, ``Simple
        Singularities and N=2 Supersymmetric Yang-Mills Theory,''
        hep-th/9411048, \PLB{344}{1995}{169};
P.C. Argyres, M.R. Plesser, and A.D. Shapere, ``The Coulomb
        Phase of N=2 Supersymmetric QCD,'' hep-th/9505100,
        \PRL{75}{1995}{1699};
A.~Hanany and Y.~Oz,
``On the quantum moduli space of vacua of N=2 supersymmetric SU(N(c)) gauge theories,''
Nucl.\ Phys.\ B {\bf 452}, 283 (1995)
[arXiv:hep-th/9505075];
%%CITATION = HEP-TH 9505075;%%
P.~C.~Argyres and A.~D.~Shapere,
``The Vacuum Structure of N=2 SuperQCD with Classical Gauge Groups,''
Nucl.\ Phys.\ B {\bf 461}, 437 (1996)
[arXiv:hep-th/9509175].
%%CITATION = HEP-TH 9509175;%%
}

\lref\DouglasXP{
M.~R.~Douglas, S.~Katz and C.~Vafa,
``Small instantons, del Pezzo surfaces and type I' theory,''
Nucl.\ Phys.\ B {\bf 497}, 155 (1997)
[arXiv:hep-th/9609071].
%%CITATION = HEP-TH 9609071;%%
}

\lref\MorrisonXF{
D.~R.~Morrison and N.~Seiberg,
``Extremal transitions and five-dimensional supersymmetric field  theories,''
Nucl.\ Phys.\ B {\bf 483}, 229 (1997)
[arXiv:hep-th/9609070].
%%CITATION = HEP-TH 9609070;%%
}

\lref\GanorXD{
O.~J.~Ganor,
``Toroidal compactification of heterotic 6D non-critical strings down to  four dimensions,''
Nucl.\ Phys.\ B {\bf 488}, 223 (1997)
[arXiv:hep-th/9608109].
%%CITATION = HEP-TH 9608109;%%
}

\lref\GanorPC{
O.~J.~Ganor, D.~R.~Morrison and N.~Seiberg,
``Branes, Calabi-Yau spaces, and toroidal compactification of the N = 1  six-dimensional E(8) theory,''
Nucl.\ Phys.\ B {\bf 487}, 93 (1997)
[arXiv:hep-th/9610251].
%%CITATION = HEP-TH 9610251;%%
}

\lref\LercheNI{
W.~Lerche, P.~Mayr and N.~P.~Warner,
``Non-critical strings, del Pezzo singularities and Seiberg-Witten  curves,''
Nucl.\ Phys.\ B {\bf 499}, 125 (1997)
[arXiv:hep-th/9612085].
%%CITATION = HEP-TH 9612085;%%
}

\lref\KazakovJI{
V.~A.~Kazakov, I.~K.~Kostov and N.~A.~Nekrasov,
``D-particles, matrix integrals and KP hierarchy,''
Nucl.\ Phys.\ B {\bf 557}, 413 (1999)
[arXiv:hep-th/9810035].
%%CITATION = HEP-TH 9810035;%%
}

\lref\DoreyPQ{
N.~Dorey, T.~J.~Hollowood and S.~P.~Kumar,
``S-duality of the Leigh-Strassler Deformation via Matrix Models,''
arXiv:hep-th/0210239.
%%CITATION = HEP-TH 0210239;%%
}

\lref\DoreyJC{
N.~Dorey, T.~J.~Hollowood, S.~P.~Kumar and A.~Sinkovics,
``Massive vacua of N = 1* theory and S-duality from matrix models,''
arXiv:hep-th/0209099.
%%CITATION = HEP-TH 0209099;%%
}

\lref\DoreyTJ{
N.~Dorey, T.~J.~Hollowood, S.~Prem Kumar and A.~Sinkovics,
``Exact superpotentials from matrix models,''
arXiv:hep-th/0209089.
%%CITATION = HEP-TH 0209089;%%
}

\lref\BerensteinSN{
D.~Berenstein,
``Quantum moduli spaces from matrix models,''
arXiv:hep-th/0210183.
%%CITATION = HEP-TH 0210183;%%
}

\lref\FujiWD{
H.~Fuji and Y.~Ookouchi,
``Comments on effective superpotentials via matrix models,''
arXiv:hep-th/0210148.
%%CITATION = HEP-TH 0210148;%%
}

\lref\FerrariJP{
F.~Ferrari,
``On exact superpotentials in confining vacua,''
arXiv:hep-th/0210135.
%%CITATION = HEP-TH 0210135;%%
}

\lref\ChekhovTF{
L.~Chekhov and A.~Mironov,
``Matrix models vs. Seiberg-Witten/Whitham theories,''
arXiv:hep-th/0209085.
%%CITATION = HEP-TH 0209085;%%
}

\lref\MarinoFK{
M.~Marino,
``Chern-Simons theory,
matrix integrals, and perturbative three-manifold  invariants,''
arXiv:hep-th/0207096.
%%CITATION = HEP-TH 0207096;%%
}

\lref\BanksNJ{
T.~Banks, M.~R.~Douglas and N.~Seiberg,
``Probing F-theory with branes,''
Phys.\ Lett.\ B {\bf 387}, 278 (1996)
[arXiv:hep-th/9605199].
%%CITATION = HEP-TH 9605199;%%
}

\lref\AharonyEN{
O.~Aharony, J.~Sonnenschein, S.~Yankielowicz and S.~Theisen,
``Field theory questions for string theory answers,''
Nucl.\ Phys.\ B {\bf 493}, 177 (1997)
[arXiv:hep-th/9611222].
%%CITATION = HEP-TH 9611222;%%
}

\lref\DouglasJS{
M.~R.~Douglas, D.~A.~Lowe and J.~H.~Schwarz,
``Probing F-theory with multiple branes,''
Phys.\ Lett.\ B {\bf 394}, 297 (1997)
[arXiv:hep-th/9612062].
%%CITATION = HEP-TH 9612062;%%
}

\lref\AharonyXZ{
O.~Aharony, A.~Fayyazuddin and J.~M.~Maldacena,
``The large N limit of N = 2,1 field theories from three-branes in
F-theory,''
JHEP {\bf 9807}, 013 (1998)
[arXiv:hep-th/9806159].
%%CITATION = HEP-TH 9806159;%%
}

%\DiFrancescoNW
\lref\DiFrancescoNW{
P.~Di Francesco, P.~Ginsparg and J.~Zinn-Justin,
``2-D Gravity and random matrices,''
Phys.\ Rept.\  {\bf 254}, 1 (1995)
[arXiv:hep-th/9306153].
%%CITATION = HEP-TH 9306153;%%
}
\lref\TanAY{C.~I.~Tan,
``Generalized Penner models and multicritical behavior,''
Phys.\ Rev.\ D {\bf 45}, 2862 (1992).
%%CITATION = PHRVA,D45,2862;
}

\lref\TanRQ{
C.~I.~Tan,
``Logarithmic Scaling Violation And Bose Condensation In One Matrix Models,''
Mod.\ Phys.\ Lett.\ A {\bf 6}, 1373 (1991).
%%CITATION = MPLAE,A6,1373;
}

%posthumous referencing:

\lref\BenaKW{
I.~Bena and R.~Roiban,
``Exact superpotentials in N = 1 theories with flavor and their matrix  model formulation,''
arXiv:hep-th/0211075.
%%CITATION = HEP-TH 0211075;%%
}

\lref\DemasureSC{
Y.~Demasure and R.~A.~Janik,
``Effective matter superpotentials from Wishart random matrices,''
arXiv:hep-th/0211082.
%%CITATION = HEP-TH 0211082;%%
}

\lref\TachikawaWK{
Y.~Tachikawa,
``Derivation of the Konishi anomaly relation from Dijkgraaf-Vafa with  (bi-)fundamental matters,''
arXiv:hep-th/0211189.
%%CITATION = HEP-TH 0211189;%%
}

\lref\FengYF{
B.~Feng and Y.~H.~He,
``Seiberg duality in matrix models. II,''
arXiv:hep-th/0211234.
%%CITATION = HEP-TH 0211234;%%
}

\lref\ArgurioHK{
R.~Argurio, V.~L.~Campos, G.~Ferretti and R.~Heise,
``Baryonic corrections to superpotentials from perturbation theory,''
arXiv:hep-th/0211249.
%%CITATION = HEP-TH 0211249;%%
}

\lref\NaculichHR{
S.~Naculich, H.~Schnitzer and N.~Wyllard,
``Matrix model approach to the N = 2 U(N) gauge theory with matter in the  fundamental representation,''
arXiv:hep-th/0211254.
%%CITATION = HEP-TH 0211254;%%
}

\lref\BenaUA{
I.~Bena, R.~Roiban and R.~Tatar,
``Baryons, boundaries and matrix models,''
arXiv:hep-th/0211271.
%%CITATION = HEP-TH 0211271;%%
}

\lref\FengZB{
B.~Feng,
``Seiberg duality in matrix model,''
arXiv:hep-th/0211202.
%%CITATION = HEP-TH 0211202;%%
}

\lref\OokouchiBE{
Y.~Ookouchi,
``N = 1 gauge theory with flavor from fluxes,''
arXiv:hep-th/0211287.
%%CITATION = HEP-TH 0211287;%%
}

\lref\christiaan{C. Hofman, ``Super Yang-Mills with flavors
from large Nf matrix models,'' hep-th/0212095.}

%\CachazoPR
\lref\CachazoPR{
F.~Cachazo and C.~Vafa,
``N = 1 and N = 2 geometry from fluxes,''
arXiv:hep-th/0206017.
%%CITATION = HEP-TH 0206017;%%
}

\def\DV{Dijkgraaf-Vafa}
\def\principalint{\int\kern-10.5pt-\kern7pt}

\Title{\vbox{\baselineskip12pt\hbox{hep-th/0211009}
\hbox{PUTP-2057}}}
{\vbox{ \centerline{Adding flavor to Dijkgraaf-Vafa}}}
\bigskip
\bigskip
\centerline{John McGreevy}
\bigskip
\centerline{{\it Department of Physics, Princeton University,
Princeton, NJ 08544}}
\bigskip
\bigskip
\noindent
We study matrix models related
via the correspondence of Dijkgraaf and Vafa
to supersymmetric gauge theories with matter in the fundamental.
As in flavorless examples, measure factors
of the matrix integral reproduce
information about R-symmetry violation
in the field theory.
The models, studied previously as models of open strings,
%similar to one studied by Kazakov,
exhibit
%(enjoy? experience?)
a large-M phase transition
as the number of flavors is varied.
This is the matrix model's manifestation of the
end of asymptotic freedom.
Using the relation to a quiver gauge theory, we extract
the effective
glueball superpotential and Seiberg-Witten curve
from the matrix model.

\bigskip
\Date{October, 2002}

\newsec{Introduction and summary}

Dijkgraaf and Vafa have found
that matrix integrals compute all of the holomorphic
information in $\CN=1$ gauge theories
with classical gauge groups
\refs{\DijkgraafFC,
\DijkgraafVW, \DijkgraafDH}.
In the interest of understanding
their proposal better, we will look at the matrix model that they prescribe
for
%\centerline{
%$\CN=2$ $SU(N)$ gauge theory with $N_f$ fundamental hypers}
%\centerline{
%mass-deformed to $\CN=1$}
%\centerline{ by a tree level
%superpotential for the adjoint chiral field $\Phi$, and masses
%for the squarks.}
%\noindent
$\CN=2$
%$SU(N)$
gauge theory with $N_f$ fundamental hypers
mass-deformed to $\CN=1$
by a tree level
superpotential for the adjoint chiral field $\Phi$, and masses
for the squarks.
These field theories, studied in \fieldtheoryrefs, exhibit
fascinating behavior as a function of the number of flavors.

Such models arise in string theory in a number of ways.
The $U(N)$ theories arise by probing the ``canonical example'' of
\DV\ with D5-branes on a noncompact
curve of the Calabi-Yau (CY) \CachazoJY.
Flavorful theories with real gauge groups arise on
$N$ D3-brane probes of D7-brane configurations
\refs{\BanksNJ,
\AharonyEN, \DouglasJS, \AharonyXZ};
this fact was recently employed
to add
holes to the BMN worldsheet \BerensteinZW.
We will instead use it to poke holes in the random surfaces
described by the \DV\ matrix integrals.

In particular, for $N=1$,
the theories with $N_f \leq 4$ (with $Sp(2N) = SU(2)$ gauge
group)
%and an extra hypermultiplet in the antisymmetric tensor representation)
arise
on a D3-brane probe of a resolved $D_4$ singularity
of F-theory.  The theory with $N_f = 4$ massless flavors
is obtained by creating the
$D_4$ singularity; in this case the
dilaton is constant and the gauge theory is scale-invariant.
This brings us to the question we would like to answer:
How is this dependence on $N_f$ manifested
in the \DV\ matrix model?

We are therefore
led to consider a matrix integral of the form
\eqn\matrixintone{
Z(g_k, m_\alpha, M) = \int d\Phi dQ d\tilde Q
\exp{\left( -W_0(\Phi) + \tilde Q_\alpha \Phi Q^\alpha  -
\sum_{\alpha=1}^{M_f}
\tilde Q_\alpha Q^\alpha m_\alpha \right)}
}
where
$$ W_0(\Phi) = \sum_{k=1}^{n+1} g_k \tr \Phi^k,$$
$\Phi $ is a complex $M\times M$ matrix,
$Q $ is a complex $M \times M_f$ rectangle,
and $\tilde Q$ is a complex $M_f \times M$ rectangle.
An arbitrary meson source
$ \tilde Q_\alpha m^\alpha_\beta Q^\beta$ can
be brought into this diagonal form by rotating the $Q$s.
As in the work of Dijkgraaf and Vafa, these are to be thought of as
line integrals over matrices of complex numbers.

Without using very much technology,
we can study in detail the
model which arises upon
integrating out the fundamentals.  This generates a $\ln (m - \Phi)$
potential for $\Phi$.   The model with this
addition to the potential can still be solved by the method
of BIPZ \BrezinSV, and actually
was studied in this way as a model for open strings
by Kazakov \KazakovCQ \foot{
Related models were also studied in \TanAY, \DistlerMT,
and in particular in \TanRQ\ where the critical behavior
at $\gamma = 2$ was explored.}.
The
coefficient of the log is $ \gamma \equiv N_f /N $, and we will show
that the cuts close up when $\gamma $ is chosen so that
the gauge theory can be conformal.  Related by supersymmetry
to this statement is the fact that the R-symmetry
becomes non-anomalous for this choice of $N_f/N$.   This
fact can be detected in the matrix model through the dependence
of the measure on $\gamma$.
Finally, we extract the glueball
superpotential and Seiberg-Witten curve
from the large-$M$ solution to the matrix model.

\subsubsec{$M$'s and $N$'s}

One of the more mysterious aspects of the
\DV\ prescription, at least to the author,
is the disjunction between the number of colors
$N$ of the gauge theory, and the number of colors $M$ of the
matrix model (which plays the role of the glueball superfield).
The addition of $N_f$ flavors to the gauge theory
only complicates this issue.  We will
%find it necessary to
introduce a number of flavors $M_f$ in the matrix model
which is again not the same as the corresponding
number in the gauge theory.
We will, however, identify the ratio
$$ \gamma \equiv {N_f\over N} = {M_f \over M} ;$$
This will be the parameter of interest in our
study of the matrix model.  We take this
as part of the prescription, but
(thinking of $\gamma$
as the weight with which holes in the random
surface contribute)
one
which is again motivated
by topological string duality \GopakumarKI.
Specifically, this is the usual \DV\ limit
for a two-node quiver gauge theory
\DijkgraafVW\ which reduces to the flavorful
theory when the dynamics of
the gauged flavor group are frozen out.
It is from this perspective that we
will be able to extract the superpotential.

\subsubsec{Related work}

Matrix models with fundamentals in this context
are mentioned in a footnote in \DijkgraafFC.
They also make an appearance in the very recent \refs{\Gorsky, \Ferretti}.
Other work on understanding and extending
the \DV\ proposal includes \refs{\MarinoFK, \ChekhovTF,
\DoreyTJ, \DoreyJC, \DoreyPQ,
\FujiWD, \FerrariJP,
\BerensteinSN, \DijkgraafPP}.

\newsec{Flavorful matrix models}

To begin, we write down the matrix integral with naive couplings and
fields marked with hats:
\eqn\matrixintzero{
Z = \int d\hat\Phi d\hat Q d\hat{\tilde Q}
\exp{\left( -\hat W_0(\hat \Phi) + \hat{\tilde Q}_\alpha
\hat \Phi \hat Q^\alpha  -
\sum_{\alpha=1}^{M_f}
\hat{\tilde Q}_\alpha \hat Q^\alpha \hat m_\alpha \right)}
}
with $$\hat W_0(\hat \Phi) \equiv \hat g_1 \tr \hat \Phi +
\hat g_2 \tr \hat \Phi^2 + \dots.$$
$\alpha$ is a flavor index; color indices will be denoted $a,b, \dots$
These hatted fields and couplings will be related to those which should have
finite large-$M$ limits (which lack hats) by an
$M$-dependent rescaling.   These hatless variables
are chosen below so that there is a well-peaked
saddle point of the $\Phi$ integral, the location of which is $M$-independent.
We will observe that the precious
$$ \half M^2 \ln M$$
term in the matrix model free energy, which
is derived from the inverse volume of the matrix model
gauge group, can also be detected by such
a propitious field rescaling.

\subsec{Integrating out the flavor}

For the moment, we are interested in the regime of couplings
where the quark masses, $m_\alpha$, are much bigger than the
bare mass of the adjoint in $W_0(\Phi)$.
In this regime, we integrate out the fast $Q$ modes at fixed $\Phi$
to get an effective potential for $\Phi$.  The integrals
over $Q_\alpha$ are $M_f$ independent gaussian integrals.
This gives
\eqn\matrixtwo{
\eqalign{
Z &= \int d \hat \Phi
e^{ -\hat W_0(\hat \Phi) }
\prod_{\alpha=1}^{M_f} \det_{ab}{}^{-1}
\left( \hat \Phi^a_b - \hat m_\alpha \delta^a_b
\right) \cr
&= \int d \hat \Phi \exp{\left(
- \hat W_0(\hat \Phi)
- \sum_{\alpha=1}^{M_f} {\rm tr} \ln ( \hat \Phi - \hat m_\alpha 1 )
\right)}\cr
}}
%Introducing $\gamma \equiv M_f/M$, and
Setting all of the masses equal
to $m$ for simplicity, this is
\eqn\matrixthree{
Z = \int d\hat \Phi ~\exp{ \left( -\hat W_0(\hat \Phi) - M_f {\rm tr}
\ln (\hat \Phi - \hat m) \right)}
}

%NOTE THAT
%the difference between $m-\Phi$ vs $\Phi - m$ is a term $\sim M \gamma$
%in the action.

A matrix integral
very similar to \matrixthree\ was studied by \refs{\KazakovCQ, \TanAY,
\TanRQ, \DistlerMT} as a
discretization of an open string worldsheet.
In this model,
the counterpart of $M_f$ is the weight accompanying a hole
insertion.
%after summing over flavors.
The logarithmic potential was chosen to reproduce
a sum over discretizations of the worldsheet boundaries,
with equal weight for arbitrary numbers of segments of the boundary.
% (insertions of $m$):
% the $1/n$ in the expansion of the $\ln$ cancels
% the symmetry factor of the diagram.
% Figure 5 on
% page 137 of Aspects of Symmetry (in ``Secret Symmetry'')

Other than a relabeling of couplings,
the difference between our model and
that of Kazakov is that
the logarithmic potential term of \KazakovCQ\ is
$$ {\rm tr} \ln ( m - \varphi^2 ).$$
This is the potential that
would arise if the $\CN=2$ superpotential were
$ \tilde Q \Phi^2 Q $ instead of $\tilde Q \Phi Q $.
The field redefinition
$\Phi = \varphi^2$ required to relate the two integrals directly
introduces a jacobian factor which adds a term
$$ \half {\rm tr} \ln \Phi $$
to the potential.  From the calculation \matrixtwo\ above we see that
this is the same as the effect of adding $M/2$ {\it massless} hypers.
We will find it convenient to solve the integral \matrixthree\ directly.
The qualitative behavior we find is the same as that found in \KazakovCQ.

\subsec{The continuum}

In order to proceed, diagonalize the matrix $\Phi$ as
$ \Phi = \CU \CD \CU^\dagger$ with
$$\CD \equiv {\rm diag}( \lambda_1, \dots, \lambda_M) ,$$
and $\CU$ unitary.
These eigenvalues are normalized as in \BrezinSV.
By the magic of logarithms, the integrand of \matrixthree\
does not depend on the angular $\CU$ variables.  Their integration
produces the Vandermonde determinant
$$ \Delta(\lambda) = \prod_{a < b} (\lambda_a - \lambda_b)^2. $$
The integral becomes
\eqn\matrixfour{
Z = \int \prod_a d\lambda_a \Delta(\lambda)
\exp{  \sum_{a=1}^M \left( -\hat W_0(\lambda_a) + \sum_{\alpha=1}^{M_f}
\ln(\lambda_a - \hat m_\alpha) \right) }
}

At this point, it is convenient to introduce a
continuum in the space of colors.
Let
$$
\lambda_a = \sqrt M \lambda(\tilde a = a/M), ~~~~
1 =  \int_0^1 d \tilde a = {1 \over M} \sum_{a=1}^M .$$
The eigenvalue density
$$\rho(\mu) =  {d \tilde a \over d \lambda }$$
is normalized to
\eqn\normalized{
\int d\mu  \rho(\mu ) = 1.
}

\vfill\eject

%see difrancesco for principal part integral symbol.

\subsubsec{Large $M$ scaling}

We now introduce the promised variables without hats:
\eqn\scaling{
\eqalign{
\hat g_2 = {g_2 \over S}, ~~~ &\hat g_3 = {g_3 \over S\sqrt M}, ~~~\dots,
\hat g_k = {g_k \over g_s M^{k/2}} = {g_k \over S M^{k/2 -1}} \cr
&\hat m_i = \sqrt M m_i \cr
\hat \Phi = \sqrt M \Phi, ~~~
&\hat Q = {1 \over M^{1/4}g_s^{1/2}} Q, ~~~~
\hat{\tilde Q} = { 1 \over M^{1/4}g_s^{1/2}} \tilde Q .
}}
Here we have finally introduced the quantity
$S \equiv g_s M $, which is fixed in the large-$M$ limit.
Note that these rescalings are closely related to those
made on dimensional grounds in the stringy realization
of the gauge theory \CachazoJY.
Plugging these into \matrixintzero,
we find that the resulting $\lambda$ integral is of the form
\eqn\fieldtheoryonaninterval{
Z = \int D\lambda(\tilde a) \exp{\left( {1 \over g_s^2 }
\CF_0[\lambda] + \CC(M) \right)}
}
with
$\CC(M)$ independent of $\lambda$
and
\eqn\largeMfreeenergy{
\eqalign{
\CF_0[\lambda] &=
S^2 \int_0^1 d\tilde a\int _0^1 d\tilde b
 \ln ( \lambda(\tilde a)  - \lambda ( \tilde b) )
- S \int_0^1 d\tilde a ~W_0(\lambda(\tilde a))
- S^2 \gamma \int_0^1 d\tilde a \ln ( \lambda(\tilde a) - m )\cr
&= S^2  \int d \lambda
\int dz \rho(\lambda) \rho (z)
 \ln ( \lambda - z)
- S  \int d \lambda \rho(\lambda) W_0(\lambda)
- S^2 \gamma  \int d\lambda \rho(\lambda) \ln ( \lambda - m )
}}
Here $W_0 (\lambda) = g_1 \lambda + g_2 \lambda^2 + \dots $.
The crucial feature of \fieldtheoryonaninterval\
is that $\CF_0[\lambda]$ is independent of $M$.
This is the normalization used by Dijkgraaf and Vafa;
the matrix action is
\eqn\normalized{
%%Z(g_k, m_\alpha, M) = \int d\Phi \int dQ \int d\tilde Q~ \exp{
- {1 \over g_s}
\left( W_0(\Phi) + \tilde Q \Phi Q - m \tilde Q Q\right)
}

Now we return to the ``constant $g$-independent term'' \BrezinSV\ $\CC(M)$.
This field- and coupling-independent
term was not relevant for previous applications of matrix integrals.
It is
\eqn\measurefactor{
\eqalign{
 \CC(M) &= {1 \over 2} M^2 \ln M - {1 \over 4 } M_f M \ln M \cr
&=(2 - \gamma )  {1 \over 4} M^2 \ln M .}}
This reproduces the leading $M$-dependence of the log of the
inverse volume of $U(M)$
\OoguriGX\ in the
field normalization we are using.
Further, it
provides the ``entropy factor'' arising from the flavor integrals\foot{
I am grateful to Nissan Itzhaki for
comments on this point.}.
The \DV\ prescription relates
the matrix model free energy to the prepotential
of the gauge theory.
In parallel with the discussion
of \DijkgraafFC\ this term leads to the following
contribution to the effective superpotential of the gauge theory:
\eqn\VenezianoYank{
W_{eff}(S)  = (2 - \gamma) N S \ln { S \over \Lambda^3},
}
up to linear terms in $S$ which are independent of the cutoff scale
$\Lambda_0$.
The prefactor of this Veneziano-Yankielowicz superpotential
is proportional to the anomaly in the $U(1)_R$ current
of the field theory.  The introduction of $N_f$ flavors in the fundamental
modifies this from $2 N$ to $(2 - \gamma) N$.
It is gratifying that this is reproduced by the simple matrix integral.

\subsec{Solution at large $M$}

In terms of the variables normalized to have a finite large-$M$ limit,
the saddle point equation is
\eqn\saddle{
{1 \over S } W_0'(\lambda) + { \gamma \over \lambda - m} =
2 \principalint d\mu { \rho(\mu) \over \lambda - \mu} .
}
Rewrite this equation as
\eqn\saddleone{
{1 \over S} W_0'(\lambda) = 2 \principalint d \mu
{\rho_0(\mu) \over \lambda - \mu}
}
where
\eqn\shifteddensity{
 \rho_0(\mu) \equiv \rho(\mu) - {\gamma \over 2} \delta ( \mu - m ),
}
or more generally in the case of arbitrary masses
$$ \rho_0(\mu) = \rho(\mu) - \gamma {1 \over 2 M_f} \sum_{\alpha=1}^{M_f} \delta(\mu - m_\alpha) .$$

Therefore the eigenvalue density $\rho_0$ satisfies the
same integral equation as that of the theory {\it without } flavors,
with the modified boundary condition
$$ \int d\mu  \rho_0(\mu) = \int d\mu \rho(\mu) - \int d\mu
\sum_{\alpha=1}^{M_f} {\gamma \over 2 M_f} \delta (\mu - m_\alpha)
= \half( 2 - \gamma). $$
This integral is to be performed over the real $\mu$ line.

For simplicity, let us consider the case of a cubic superpotential,
in the case of a single cut,
\ie\ choose the potential $W_0$ to have a single critical point.
Placing the cut at $[ 2a, 2b]$, the BIPZ solution for the
resolvent
\eqn\resolvent{
 \omega_0(\lambda) = \int_{2a}^{2b} d\mu { \rho_0(\mu)\over \mu - \lambda}
}
of the theory without flavors
determines the solution of \saddle.  This
is \BrezinSV\
\eqn\solutionforomega{
S \omega_0(z) = 2 g_2 z + 3 g_3 z^2 -
\left( 2 g_2 + 3 g_3 (a+b) + 3 g_3 z \right) \sqrt{ (z-2a)(z-2b)}.
}
The conditions determining the positions
$a,b$ of the ends of the cut are
determined by the behavior of
\solutionforomega\ at $z \to \infty$.
For the model with flavor, these are
\eqn\numberone{
3g_3 (b-a)^2 + 2 (a+b) (2 g_2 + 3 g_3 (a+b)) = 0 }
\eqn\numbertwo{
{1 \over S}  (b-a)^2 ( 2 g_2 + 6 g_3 (a+b) ) = 2 - \gamma.
}
Up to coupling redefinitions,
these differ from equations $(46)$ of \BrezinSV\
only by the replacement $2 \mapsto 2 - \gamma$
in the second condition.
Note that they do not depend on $m$.

\subsubsec{Healing of cuts}

Assume there is a stable vacuum at the origin,
and place all of the eigenvalues there,
\eg\ consider $ g_k = 0 , k\geq 3$, $g_2 >0$.
From $\numberone $ and $\numbertwo$ we immediately see
that when $\gamma \to 2$, $ b-a \to 0$.  That is, the cut closes up.  Beyond
$\gamma =2$, the cut at the origin moves into the complex plane.
In the context of a Hermitian matrix integral \KazakovCQ\ this
was interpreted as a large-$M$ phase transition beyond which
the theory lacked a stable solution.  However, when
the saddle point we are studying is that of a holomorphic line integral,
this is not so catastrophic.  In fact, the cut surrounding
the unstable extrema of $W$ always extend
into the imaginary direction of $\lambda$.

In the matrix models for confining gauge theories, the size
of the cuts which hold the eigenvalues goes like
the IR scale $\Lambda$ of the field theory.
The closing of the cuts is a signal that the theories
become scale invariant, and then no longer asymptotically free
as $N_f $ passes through $2 N$.

We can see the corresponding effect for more general
tree-level superpotentials as follows.
As a consequence of the saddle equation \saddleone,
the resolvent
$$\omega_0(x) = \int d\lambda {\rho_0(\lambda) \over x - \lambda }$$
of the $\gamma=0$ theory satisfies at large $M$
an algebraic equation of the form \refs{\eg\ \DiFrancescoNW}
\eqn\loopeqn{
\omega_0(x)^2 +
{ 1 \over S} \omega_0(x) W'_0(x) + {1 \over  4 S^2} f_0(x) = 0.
}
Here
%$S = g_s M$ is the 't Hooft coupling/glueball field,
$f_0(x)$ is a degree $n-1$ polynomial in $x$, which
should be thought of as a function of $S_i = g_s M_i$,
with $M_i$ the number of eigenvalues in the $i$th cut.
This polynomial $f_0$ differs from
the one in the $\gamma = 0 $ solution
through its dependence on the locations of the cuts.
Note that the resolvent of the theory with flavor is
related to $\omega_0$ by the addition of a pole term
\eqn\wehavepoles{ \omega(x) = \omega_0(x) + {\gamma \over x - m} .}
The remainder term, $f_0$, can be written \DiFrancescoNW\
$$
f_0(z) = 4 S
\int dw ~\rho_0(w) { W_0^{\prime} (w) - W_0^{\prime}(z)
 \over w - z};
$$
this is a polynomial of degree $n-1$ in $z$,
$ f_0(z) = \sum_{k = 1}^{n-1} b_k z^k $.
The fact that $\int dw~ \rho_0(w) = 2 - \gamma$ then implies that
as $\gamma \to 2$, the leading term, $b_{n-1}$, vanishes.
In more detail,
\loopeqn\ is a quadratic equation for $\omega_0$, which therefore
has the solution
$$ 2 S \omega_0(z) = W_0^\prime(z) \pm \sqrt{ ( W_0^\prime(z))^2 - f_0(z) } .$$
From this expression
we learn that
$$b_{n-1} = - 2 (2 - \gamma) (n+1) g_{n+1} S .$$
But, we also know
from \CachazoJY\ and from \OokouchiBE\ that
$$ b_{n-1} = - 4 (n+1) g_{n+1} { \del W_{eff} \over \del \ln \Lambda^{2N}} .$$
Therefore, we see from the matrix model
that the superpotential is independent of
the IR scale of the gauge theory when $\gamma \to 2$.
If $m=0$, all the cuts are healed, in accord with the
fact that in that case
scale invariance is never broken.

%\subsubsec{Real gauge groups}

%For field theories with gauge group
%$SO(2N)$, the corresponding
%matrix model differs in that $\Phi$ is a real matrix.
%\MarinoFK.
%This changes the Vandermonde determinant to
%$$ \Delta = \prod_{a<b} | \lambda_a - \lambda_b | .$$
%%Similarly, the measure factor is $ {1 \over 4} M^2 \ln M $.
%This leads to the replacement of $2 - \gamma$ in the formulas above
%by $1 - \gamma$, which correctly reproduces the beta function and
%R-symmetry anomaly of the $SO$ gauge theories.

\subsec{The Seiberg-Witten curve and glueball superpotential}

In this subsection we will extract
the glueball superpotential from the
free energy of the matrix model, and
the Seiberg-Witten (SW) curve
of the gauge theory
from the loop equation\foot{Please
note that this subsection was rewritten for version three,
after the appearance of \refs{
\BenaKW,\DemasureSC,\TachikawaWK,\FengYF,\FengZB, \ArgurioHK, \NaculichHR,
\BenaUA} in which superpotentials
were found using matrix models with a finite number of flavors.}.

In the canonical example of \DV, the free energy
of the matrix model determines the effective
glueball superpotential of the field theory via
\eqn\weff{
 W_{eff}(S_i) = \sum_i N_i {\del \CF_0\over \del S_i}
- 2 \pi i \tau S  }
with $\CF_0(S_i) = g_s^2 \ln Z$.
%with $S_i = g_s M_i$; $M_i$ is the number of eigenvalues
%in the $i$th cut, and
Thinking of the flavors as arising from the $\CN=2$ quiver with two
nodes of respective ranks $N$ and $N_f$ in the limit that the
coupling of the flavor group vanishes,
this formula is modified in our case to\foot{While this
paper was being revised,
a closely related observation was made independently in \christiaan.}
\eqn\weffhere{
 W_{eff} = - 2 \pi i \tau S
+ \sum_i N_i {\del \CF_0 \over \del S_i}
+ N_f {\del \CF_0\over \del S_f}
 }
where $S_f = g_s M_f$.
Because the flavor gauge coupling is zero, we do not
include the dynamics of the corresponding matrix variables,
and treat them as a background.
\weffhere\ has been written for
the case where all of the flavor branes are
coincident (\ie\ the masses for the flavors
are identical), but the generalization is clear.

A convenient way \NaculichHR\
to compute $\del \CF_0 \over \del S_i$ is
by varying the eigenvalue density according to
$$ \rho(z)  \mapsto \rho(z) + \delta S_i {1 \over S} \delta( z - e_i) $$
with $e_i$ some fixed point inside the $i$th cut.
From \largeMfreeenergy\ evaluated in the saddle, we find
$$ {\del \CF_0 \over \del S_i } =
\int_{e_i}^{\Lambda_0} dx
\left[ W_0^\prime(x) -
 S \left( 2\int d\lambda {\rho(\lambda) \over x-\lambda} -
{\gamma \over x-\lambda}
\right)\right]_{\rho(\lambda) = \rho_0(\lambda) +
{\gamma\over 2} \delta(\lambda -m)}
$$
We have ignored the additive constant, and
$\Lambda_0$-independent terms linear in $S$,
which can be absorbed in the bare coupling, $\tau$.
This expression
%for $\del \CF_0 \over \del S_i $
is identical to that for the case without flavor;
the contributions proportional to $\gamma$ cancel
each other.
So we have
\eqn\sphere{
 {\del \CF_0 \over \del S_i} =
\int_{B_i} dx W_0^\prime(x)
-  2 S \int dw \rho_0(w) \int_{B_i} {dx \over x -w }
= \int_{B_i} dx \left( W_0^\prime(x) + 2 S \omega_0(x) \right)
}
where $B_i$ is a contour running from the reference point $e_i$
to the cutoff $\Lambda_0$.
As in the case without flavor,
we introduce $y$, the singular piece of the
($\gamma = 0$) resolvent, by completing the square in \loopeqn\ \DijkgraafFC\
$$y(x)  = 2 S \omega_0(x) + W_0'(x) .$$
Then \sphere\ becomes
\eqn\finalsphere{
 {\del \CF_0 \over \del S_i} =  \int_{B_i} y dx .
}

Similarly, we can determine the value of the new term
in \weffhere\ by varying
\eqn\varygamma{
 \rho(z) \mapsto \rho(z) + {\delta \gamma \over 2} ~\delta (z - m_0) ;}
$m_0$ is the mass at which we add the new flavor.
The factor of two in \varygamma\ appears because of the corresponding
factor in \shifteddensity.
Under this change,
\eqn\disc{
%\eqalign{
\delta_\gamma  \CF_0 =
\half \left[ - S  W_0 (m_0)
+ 2S^2 \int d\lambda ~\rho(\lambda) \ln ( \lambda - m_0)
- S^2  \gamma \ln (m_0 -m)
\right]_{\rho(\lambda) = \rho_0(\lambda) +
{\gamma \over 2}\delta (\lambda - m)}.
%\cr
%&- { S\over \gamma} \sum_{\alpha=1}^{M_f} \int d\lambda ~\rho(\lambda)
% \ln(\lambda - m_i)
%- {S \over \gamma} \sum_{\alpha=1}^{M_f} \rho(m_0) \ln(m_0 - m_i) .
%}
}
%Note that $\rho_0$ in the saddle-point solution
%\shifteddensity\
%depends on $\gamma$ through its dependence on the locations
%of the cuts (Eqn. \numbertwo).
%The last term in \disc\ arises from the measure factor
%\measurefactor.
%so this is not merely
%$\CF_0(S, \gamma =0) + \dots$.
%We note that evaluating the $\delta$ functions gives
%a divergence of the form ${\gamma^2\over 2} \ln (m-m)$, which
%appears to be independent of the $S_i$.
Again up to irrelevant terms, this can be rewritten to give
$$
\eqalign{
{ \del \CF_0 \over \del S_f} &=
{1 \over 2} \left[ - \int_{m_0}^{\Lambda_0} dx W_0^\prime(x)
+ 2 S  \int_{m_0}^{\Lambda_0} dx
\int {d \lambda \over \lambda - x}
\left( \rho_0(\lambda) + {\gamma\over 2} \delta (\lambda - m) \right)
+ \gamma S \int_{m_0}^{\Lambda_0} {dx \over x -m} \right]\cr
&=
-{1  \over 2} \int_{m_0}^{\Lambda_0} dx \left(
2 S \omega_0(x) + W_0^\prime(x) \right) =
- \half \int_{m_0}^{\Lambda_0} ydx.
}
$$

We therefore find
\eqn\finally{
W_{eff}(S_i) =  \sum_i N_i \int_{B_i} y dx
%\half N_f \int_{m \Lambda_0^{1/2}}^{\Lambda_0^{3/2}} ydx
- \half N_f \int_{m }^{\Lambda_0} ydx
}
with $y$ defined by
$$y^2  = (W_0')^2 - f_0 $$
%$ y^2 = P_N^2(x) + f_0(x) ,$ and
Here, the coefficients of $f_0(x) = \sum_k b_k(S_i) x^k $ are determined by
the dynamical glueball fields $S_i$ according to
$$S_i = \int_{A_i} y dx$$
where $A_i$ is a contour encircling the $i$th cut.
This is the formula for the effective superpotential
predicted \CachazoJY\ from
the physics of D5-branes on the generalized conifold.

\lref\Freddy{F. Cachazo, private communication.}

Taking $N_i=1$, minimization of this superpotential with respect to
variations of the polynomial $f_0$ has been shown
\refs{\Freddy, \NaculichHR, \OokouchiBE}
using the methods of \CachazoPR\
to result in the correct Seiberg-Witten curve
\eqn\sw{ y^2
= \prod_{a=1}^N (x - \phi_a)^2 + \Lambda^{2N - N_f} ( x - m)^{N_f},
}
where $\phi_a$ are the critical points of $W_0$.

\newsec{Discussion and prospects}

In this paper, we have focused on the regime of couplings
$ m_\Phi << m_Q $
(though of course we can still perform the gaussian integral
over $Q$ in the other regime)
where the interesting observables involve the
adjoint field, as in the case without flavors.
It will be interesting to try to compute other observables involving
light quarks.  In the $SU(N)$ version
of these theories, these include
baryon operators
$$
\eqalign{
&Z[M,B,\tilde B] = \int d\Phi dQ d\tilde Q
\exp{\left(
W_0(\Phi) - \tilde Q_\alpha \Phi Q^\alpha  + \sum_\alpha
\tilde Q_\alpha Q^\alpha m_\alpha\right)} \cr
&\exp{\left( \tilde Q_{\alpha}^ a M^\alpha_\beta Q^\beta_a
+ B^{\alpha_1 \dots \alpha_N}
\epsilon_{a_1 \dots a_N} Q_{a_1}^{ \alpha_1} \dots Q^{a_N}_{ \alpha_N}
+ \tilde B^{\alpha_1 \dots \alpha_N} \epsilon_{a_1 \dots a_N}
\tilde Q^{a_1}_{ \alpha_1} \dots \tilde Q^{a_N}_{ \alpha_N} \right)}}
$$
These baryon sources exist for $N_f \ge N$, corresponding to a value
of $\gamma$ at which we have
not yet detected any change in behavior of the matrix model.

%Consider for the moment just a color-singlet meson source:
%\eqn\mesonsone{
%%Z(M) = \int d\Phi dQ  d\tilde Q
%e^{ W_0(\Phi) - \tilde Q_\alpha \Phi Q^\alpha }
%\exp{\tilde Q_{\alpha a} M^\alpha_\beta Q^{\beta a}  }.
%}
%Because $M$ is gauge invariant, we can still do the $Q$ integral, and obtain
%\eqn\mesonstwo{
%Z(M) = \int d\Phi  { e^{W_0(\Phi)}\over \det{\left(
%\Phi \otimes 1 - 1 \otimes M \right) } }
%}

%One can take the point of view that the flavors arise from an $\CN = 2$
%quiver with two nodes
%$N$ and $N_f$ where
%we turn off the coupling of the flavor gauge group.
%Then $M$ arises from the adjoint chiral field in the $\CN=2$ vectormultiplet
%for the flavor group.  In this form, the integral \mesonstwo\ was studied in
%the recent \DijkgraafPP.
%From this perspective, the convenience of
%making a continuum of the space of flavors
%is less mysterious.

\subsubsec{Beyond the transition}

Consider the F-theory realization of the related
symplectic models mentioned in the introduction
($Sp(N)$ with $N_f \leq 4$ and an antisymmetric tensor).
The flavor symmetry (which is $SO(8)$ in the
critical case) is the gauge symmetry on the D7-branes.
Increasing $N_f/N$ beyond the critical value is achieved by
adding more D7-branes to the $D_4$ singularity.
This is possible without destroying the triviality
of the canonical bundle, and one obtains in this
way collections of D7-branes with
the exceptional series of gauge groups
of rank up to 8.  The D3 probe theories are then
field theories with exceptional flavor symmetry
\refs{\eg\ \MorrisonXF, \DouglasXP, \GanorXD, \GanorPC, \LercheNI}.

It has become clear that
the complex $x$-plane of $\Phi$ eigenvalues can be
identified with
the image of a fibration of a noncompact
CY geometry.
A cut which holds the eigenvalues
in the large $M$ solution is identified
with the image in the $x$-plane of a three-cycle
in this geometry (after a geometric
transition induced by the flux
generating $W_0(\Phi)$).
It is therefore
tempting to speculate that tuning $\gamma$ past
the critical value is the matrix model
version of performing an extremal
transition in the CY geometry,
during which the three-cycle shrinks
and one finds an even-dimensional cycle
which can be resolved.
The fact that a shrinking del Pezzo
four-cycle in a CY realizes a theory
with exceptional flavor symmetry
leads to a clear candidate for the nature
of this new direction.

\vfill\eject
\subsubsec{Some remaining issues}

\item{1.} Our calculations should extend to the
case of real gauge groups, and in particular to the
theories with extra tensor representations,
arising from D3-brane probes of F-theory.

%\item{2.} We have not yet succeeded in extracting the $S_i$-dependence
%of the matrix model free energy $\CF_0(g_k, m, \gamma)$,
%nor in reproducing the expected SW curve \sw.

\item{2.} ``$uv$'' completions of the Seiberg-Witten curve can
be seen from the matrix model.
As explained in \BerensteinSN, including more of the
``fractional branes'' of the CY singularity
allows one to determine an embedding of the SW curve
in a threefold of a form such as
$$ uv = F(x,y).$$
This is important, for example, because it
will allow
one to identify
the resolution involved in the extremal transition
proposed above.

\item{3.}
The field theories obtained on 3-brane probes of F-theory
exhibit S-duality.
In a beautiful series of papers
\refs{\DoreyTJ, \DoreyJC, \DoreyPQ, \DijkgraafPP},
the S-duality of the $\CN=4$ theory and its $\CN=1$
deformations
has been found via the solution of the corresponding
matrix integral in \KazakovJI.
The symplectic matrix model with the corresponding matter content
should also have modular behavior in $\tau$; it
shares the feature with the $\CN=4$ theory that
the path integral over the matter cancels
the Vandermonde for the adjoint matrix.

\item{4.} Kazakov \KazakovCQ\ computes ``average numbers of holes''
and ``average lengths of holes'' in the random surfaces,
from the large $M$ solution to the matrix integral.
These observables exhibit more detailed critical behavior
than we have discussed thus far as $\gamma, m, g_k$ are varied.
The transition to
``torn surfaces'' with large holes likely has an interpretation
in terms of the appearance of a Higgs branch in the gauge theory
when an $m_\alpha$ approaches an eigenvalue of $\Phi$.
It will be interesting to find a superpotential
via which we can fix the moduli at a point on this
Higgs branch.

\bigskip

\centerline{\bf{Acknowledgements}}

I am especially grateful to Igor Klebanov
for several extremely helpful discussions.
I would also like to thank
Ofer Aharony, David Berenstein, Freddy Cachazo,
Nissan Itzhaki, Kostas Skenderis, and Herman Verlinde for useful comments,
and Christiaan Hofman for correspondence on the revision.
This work was supported by a Princeton University Dicke Fellowship.

\listrefs

\end